\documentclass{emulateapj}

\usepackage{color}

\slugcomment{To appear in ApJL}

\shorttitle{Rescuing the IMF}
\shortauthors{Kotulla, Fritze \& Gallagher}

\begin{document}


\title{Rescuing the Initial Mass Function for Arp~78}

\author{Ralf Kotulla and Uta Fritze}
\affil{Centre for Astrophysics Research, 
University of Hertfordshire, College Lane, Hatfield AL10 9AB, United
  Kingdom} \email{r.kotulla@herts.ac.uk, u.fritze@herts.ac.uk}
\and
\author{John S Gallagher III}
\affil{Department of Astronomy, University of Wisconsin, 475 N. Charter Street, Madison, WI 53706, USA}
\email{jsg@astro.wisc.edu}


\begin{abstract}
  We present deep R and narrow-band ${\rm H\alpha}$ images of Arp~78 obtained
  with the WIYN 3.5-m telescope on Kitt Peak. GALEX observations had shown a
  very extended UV structure for this system, reaching beyond the optical
  radius of Arp~78 and also beyond its previously known ${\rm H\alpha-}$
  radius. Our new ${\rm H\alpha}$ data now show agreement not only with the
  spatial extent of the near- and far-UV maps, but also in terms of structural
  details. Star formation rates derived from L(H$\alpha$) and L(FUV) are in
  reasonable agreement, indicating that in this case the upper stellar IMF in
  the UV-bright outer arm is relatively normal.  The star forming sites in the
  outer arms are younger than $\sim 15\,$Myr and massive enough to properly
  sample the IMF up to high masses; their low optical visibility
    evidently is a property of their youth.
\end{abstract}

\keywords{galaxies: interactions --- galaxies: stellar content --- galaxies: individual (Arp~78) --- ultraviolet: galaxies}

\section{Introduction}
A major surprise from early data taken with the GALEX ultraviolet explorer
satellite was the detection of extended near- and far-UV (NUV, FUV) emission
in the extreme outer environment of star-forming galaxies like M83 and NGC
4625 \citep{Thilker+05,GildePaz+05}. Other examples include both apparently
undisturbed spirals where the UV emission reveals the inside-out growth of the
stellar disk and galaxies that are surrounded by filaments and substructure
indicative of tidal encounters with active star formation (SF) going on within
these filaments.
 
An analysis 189 disk galaxies (types S0-Sm) within 40~Mpc from the GALEX Atlas
of Nearby Galaxies \citep{GildePaz+07,Thilker+07} establishes that extended UV
(XUV) galaxies are surprisingly common, showing up in $>30 \%$ of this sample.
Two classes of XUV galaxies are defined from this sample.  Arp~78 (NGC~772),
that we study here, belongs to the XUV-type 1 class of objects, that make up
$\geq 20 \%$ of the 40~Mpc sample and of which over 75\% show optical or HI
morphological evidence for recent interactions or external perturbations. A
prototype of this class is M83.

These galaxies have structured, UV-bright, optically faint emission features
beyond their normal optical radii and in regions beyond the traditional SF
threshold.  The latter is defined as the surface brightness contour
corresponding to $\Sigma_{\rm SFR} = 3 \times 10^{-4}\,{\rm
  M_{\odot}\,yr^{-1}\,kpc^{-2}}$, evaluated at 1 kpc resolution. With
\cite{Kennicutt98}'s star formation rate (SFR) calibration, this threshold
corresponds to $\mu_{\rm FUV} =27.25$ ABmag/arcsec$^2$ or $\mu_{\rm NUV}
=27.35$ ABmag/arcsec$^2$. These UV surface brightness thresholds correspond to
an HI column density threshold for actively star-forming zones, as predicted
by \cite[see also \citealt{Thilker+07}]{Schaye04}, as well as to the
H${\alpha}$ ``edge'' in galaxies from \cite{MartinKennicutt01}'s sample, as
demonstrated by \cite{Boissier+07}.

A great deal of discussion has focused on objects where UV emission was
detected at larger radii than H${\alpha}$ emission from HII regions, as was
the case for Arp~78.  A variety of possible explanations for this discrepancy
have been put forward, e.g. a top-light IMF as a consequence of a low level of
SF with the resulting low star cluster masses precluding the formation of
ionizing stars while still forming enough stars below the ionization limit to
account for the FUV flux \citep{WeidnerKroupa06,Boissier+07}. Wholesale
truncations of the IMF at the upper end also have been suggested to simply cut
off sources of photoionization.  This situation could be attributed to the low
gas column densities in the outskirts of galaxies. Under these conditions
fragmentation of a molecular cloud happens too quickly for it to grow to a
point where it can readily produce stars massive enough to ionize the
surrounding gas \citep{KrumholzMcKee08}.

Alternatively, age effects have been discussed, in the sense that presumably
episodic SF events ceased in the outskirts of some galaxies long enough in the
past for HII regions to have faded below detection, but recently enough for
the stellar UV flux to still be measurable \citep{ZaritskyChristlein07}.
Leakage of ionizing photons in the very low density environment of these outer
regions provides yet another possible explanation, as this would act to
decrease the observability of photoionized regions \citep{OeyKennicutt97}.

In some undisturbed face-on spirals, \cite{Ferguson+98b} were the first to show
that very deep H$\alpha$ exposures revealed small and low-luminosity HII
regions beyond -- and sometimes {\em far} beyond -- the optical radius, giving
evidence for low-level SF activity going on beyond the optical radius and
interpreted as a signature of the inside-out growth of stellar disks. Another
factor then is the question of detectability of HII regions versus the UV
light of stars.

Here we present deep narrow band H${\alpha}$ imaging of Arp~78, including its
outer regions, to check whether we can find H${\alpha}$ flux from HII
counterparts to the XUV flux detected by GALEX.  Our approach involves
comparing SFRs derived from L(H$\alpha$) with those based on L(FUV) assuming a
normal relationship between these two measures of massive stellar populations.
In this way we can check whether or not the outer regions of Arp~78 have
normal young stellar populations.

\section{Arp~78 -- NGC~772}

Arp~78 is a luminous (${\rm M_B = -21.6}$) galaxy with pronounced spiral
structure at an adopted distance of ${\rm 34~Mpc}$ ($\rm v_r =
2472\,km\,s^{-1}$). It is experiencing multiple interactions involving its
spectroscopically confirmed low-luminosity (${\rm M_B = -18.2}$) elliptical
satellite, NGC~770, at a projected distance of $\sim 30$ kpc, plus two more
companions (${\rm M_B = -15.5}$ and $-16.2$, respectively) within projected
distances $\sim 400~$kpc \citep{Zaritsky+97}.

The outer regions of Arp~78 feature an obviously disturbed, one-sided spiral
arm-like morphology that appears to be a tidally driven structure. The XUV
emission in Arp~78 is clearly associated with this region, as seen in Fig.
\ref{fig:fuvha}. Whether or not the stellar population age in the filaments
agrees with the nuclear starburst age of $\sim 2$ Gyr as determined by
\cite{Ganda+07} and possibly even with the stellar population age of $3 \pm
0.5$ Gyr given by \cite{Geha+05} for the counter-rotating disk in the
companion NGC~770, are subjects of our ongoing more extensive multi-band
analysis.

\section{Observations and Data reduction}
Our optical observations were obtained using the WIYN\footnote{The WIYN
  Observatory is a joint facility of the University of Wisconsin-Madison,
  Indiana University, Yale University, and the National Optical Astronomy
  Observatories.} 3.5~m telescope at Kitt Peak equipped with the MiniMosaic
camera. MiniMo consists of two 2K$\times$4K CCD chips with a spatial sampling
of $0.14$ arcsec per pixel resulting in a field-of-view of $9.5\times9.5$
arcmin. The seeing in the R-band was $\approx1.0''$ and $\approx 1.2''$ in
H$\alpha$.

Data reduction consisted of overscan- and bias-subtraction, flat-fielding and
cosmic ray removal. We took special care to correct for slight gain variations
between the two CCDs to yield a flat background across the full FoV by
multiplying them individually with correction factors $\approx 1$ until
sky-noise and background level were identical in all readout zones. The
resulting frames were then aligned by matching the positions of several stars
in each frame and stacked; bad-pixel-masks were used to remove bad pixels and
the small gap between the detectors.

Obtaining a proper continuum subtraction requires that we match the point
spread function (PSF) in the R and H$\alpha$ filters. We accomplished this by
iteratively smoothing the R-band image with a Gaussian of varying widths in
the x- and y-directions until the PSFs in both filters matched and residuals
were acceptable.  To remove the continuum contribution from the H$\alpha$
image, we measured the intensity of several stars in both the R-band and
H$\alpha$ frame, scaled the R-band so that the stars have on average the same
count rates in both R and H$\alpha$ and subtracted this frame from the
H$\alpha$ narrowband exposure.  The resulting scaling factor is in good
agreement with that derived from the widths of the filter transmission curves,
but allows for a more accurate continuum subtraction. Although the WIYN W16
narrow band filter also includes the [NII] emission lines, we refer to the
continuum subtracted data as H$\alpha$ images since this is the dominant
source of emission line flux.

The GALEX images have been obtained from the GALEX science archive at the
Space Telescope Science Institute. To minimize offsets introduced by slightly
differing coordinate systems we aligned the FUV and NUV frames relative to the
optical data by matching the positions of several stars. The full width at
half maximum of the FUV PSF is $4.2''$ sampled with $1.5''$ pixels
\citep{Morrissey+07}.

\section{Results and Implications for IMF}
\begin{figure*}
\includegraphics[width=\textwidth]{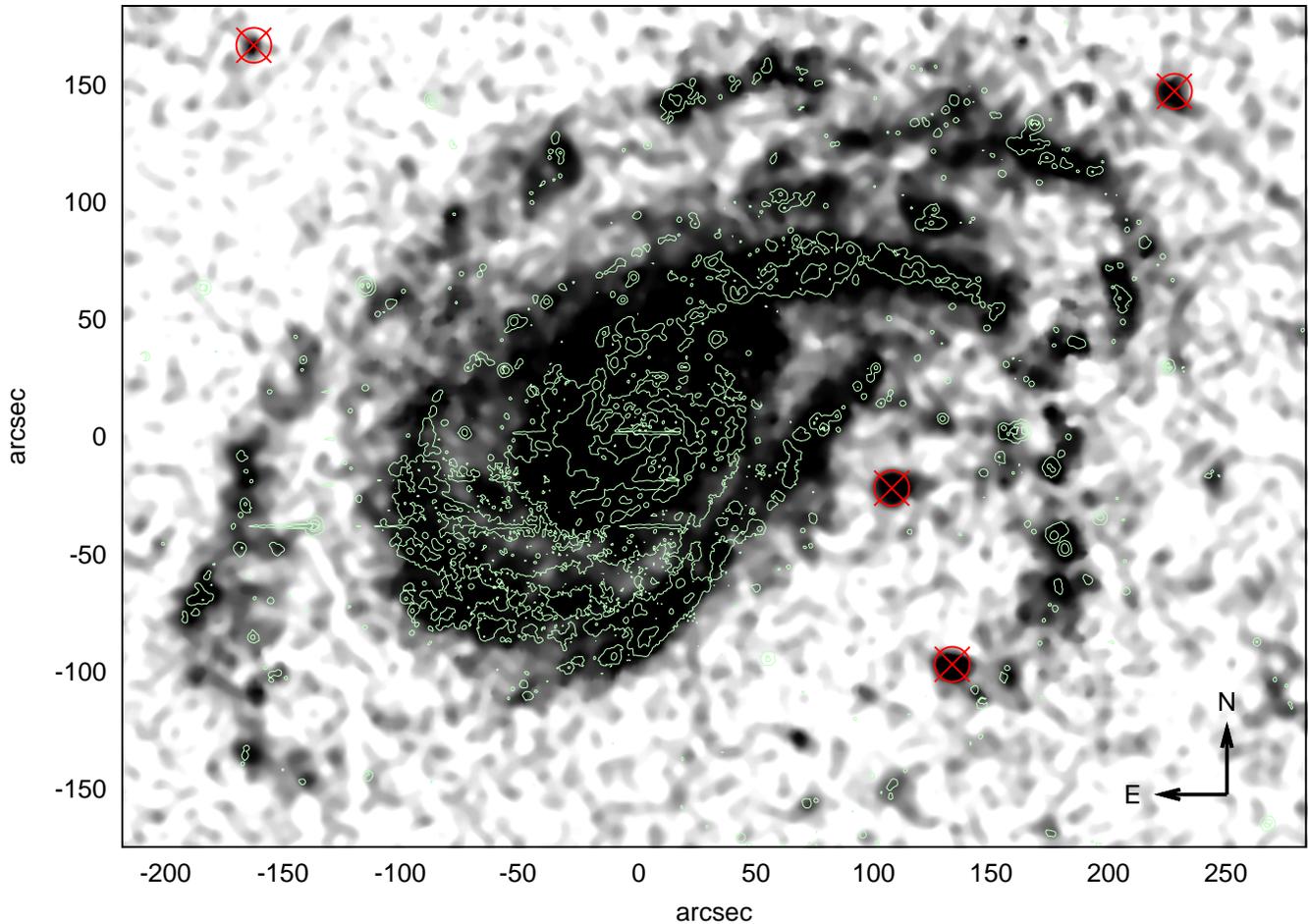}
\caption{The far-UV image of Arp~78 shown as grayscales is overplotted with
  contour lines for the H$\alpha$ emission. The four circles mark positions of
  three background galaxies and a point source (probably an AGN or a hot star)
  that show up in the FUV, but not in our H$\alpha$ images.}
\label{fig:fuvha}
\end{figure*}

Figure \ref{fig:fuvha} shows the FUV image of Arp~78 as grayscale with WIYN
narrow band H$\alpha$-brightness contours overplotted. To suppress noise from
the image we used adaptive and median-filtering of the FUV image. The
H$\alpha$ contours, however, have been constructed from the original continuum
subtracted H$\alpha$ frame to make sure this filtering does not induce
artifacts into our results.

This Figure clearly shows a nearly perfect match between the FUV and HII
regions, as is expected if both the UV-flux and H$\alpha$ flux are emitted by
the young stars in the same regions where the IMF extends up to high stellar
masses. The close coincidence between FUV- and H$\alpha$ emitting regions both
in the inner disk and in the outer tidal features is {\em inconsistent} with a
scenario where a low level of star formation in the XUV-disk leads to a low
upper stellar mass cutoff of the IMF. It also excludes timing models where the
FUV-emitting star clusters are systematically too old to produce significant
H$\alpha$ emission through photoionization, i.e. older than $\rm\approx
7\,Myr$.

Several small isolated regions seen in the UV GALEX images but lacking
H$\alpha$ emission can be identified as background galaxies in our WIYN
optical images; they are marked by encircled crosses in Figure~1. 
There are also a few horizontal structures visible in the H$\alpha$
contour map: these originate in chip defects or saturated stars that could not
perfectly be corrected for in the data reduction process.

\subsection{Comparing SFRs from FUV/NUV and H$\alpha$}
The existence of HII regions could be consistent with models where the upper
IMF is biased in outer regions of galaxies, e.g. the \cite{WeidnerKroupa06}
models.  To test for this possibility we compare the SFRs obtained from the
FUV and H$\alpha$ luminosities using \cite{Kennicutt98}'s calibrations.
Although those were obtained -- and hence are strictly accurate only -- for
close-to-solar metallicity, they provide a reasonable first estimate for the
outer regions of Arp~78. While central spectroscopy has revealed solar to
slightly super-solar abundances \citep{Ganda+07}, we expect the outer regions
to have subsolar -- albeit probably not dramatically subsolar -- abundances.
If IMF biasing exists in the XUV regions, then we expect the SFR derived from
the H$\alpha$ luminosity to be significantly lower than that based on FUV
data.

\begin{figure}
  \includegraphics[width=\columnwidth]{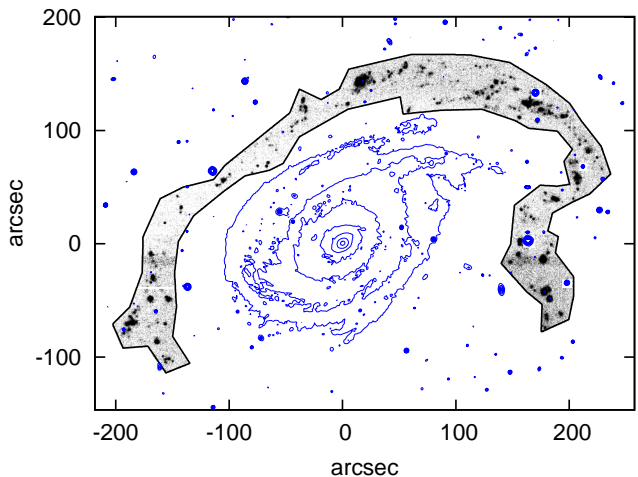}
  \caption{Grayscale image of the region of our continuum subtracted H$\alpha$
    frame used to compute the SFR (bordered by the solid black line). For
    reference we also show contour lines of the R-band continuum image with
    thin blue lines.}
  \label{fig:haregions}
\end{figure}

To obtain the fluxes in the outer tidal filaments, we masked the outer
ring-like structure (cf. Fig. \ref{fig:haregions}) and obtained integrated
count rates for this region from which we derive fluxes.  Using the Galactic
extinction law from \cite{Valencic+04}, we derive a value of $\rm
A_{FUV}=0.63\,mag$ at 1540 {\AA} from the Galactic extinction $\rm
E(B-V)=0.073\,mag$ towards Arp~78 \citep{Schlegel+98}. At the distance of $\rm
34\,Mpc$ towards Arp~78, this yields a FUV-luminosity of $\rm
2.5\times10^{39}\,erg\,s^{-1}\,cm^{-2}\,\AA^{-1}$. Based on uncertainties in
the background determination and sky- and readout noise we estimate that this
luminosity is accurate to $\approx \pm15\%$. From this FUV luminosity we
estimate a SFR of $\rm SFR_{FUV}=(0.3\pm0.05)\,M_{\sun}\,yr^{-1}$ using the
calibration from \cite{Kennicutt98}.

To derive an H$\alpha$ flux we use the calibration data from the R-band to
convert the count rates we measure into physical units.  For the photometry in
the region identical with the one used to derive the FUV flux we considered
two extreme cases: To derive an upper limit to the SFR we only count pixels
with counts $> 1 \sigma$ above the background level as determined from the
nearby sky. Using an effective filter width of $\rm 62\,\AA$ for the WIYN-W16
filter yields $\rm L(H\alpha) = 3.2\,\times 10^{40}\,erg\,s^{-1}$ for the
regions marked in Fig.  \ref{fig:haregions}, corresponding to a SFR in the
tidal debris region of $\rm SFR_{\rm H\alpha} = 0.24\,M_{\sun}\,yr^{-1}$,
also based on the \cite{Kennicutt98} calibration. A 3$\sigma$ threshold, taken
to be a lower limit to the true SFR, reduced the SFR$_{\rm H\alpha}$ to $\rm
0.15\,M_{\sun}\,yr^{-1}$.  The uncertainty is dominated by potential
systematic errors, such as imperfect registration of the FUV and H$\alpha$
frames, variations of the continuum scaling factor (on the order of $10\,\%$
or $\rm 0.02\,M_{\odot}\,yr^{-1}$)
and large scale fluctuations in the background subtraction due to remaining
flat-field errors ($\rm\approx 0.05\,M_{\odot}\,yr^{-1}$ if integrated over the
masked area).  We therefore adopt a SFR$_{\rm H\alpha} =$ 0.2 $\pm$0.05
(random) $\pm$0.1 (systematic)~$\rm M_{\odot}\,yr^{-1}$ .  Thus the ratio of
FUV-to-H$\alpha$ SFRs lies in the range of 0.5-1.2, close to expectations for 
a normal upper stellar mass function. 

The observed ratio of SFRs depends on a variety of factors other than the
present day average upper stellar mass distribution. FUV luminosities can be
reduced by dust absorption internal to Arp~78. The L(H$\alpha$) is very
dependent on detecting faint emission against a comparatively bright optical
sky background. This problem can be especially serious in low surface
brightness regions where 30-50\% of the H$\alpha$ flux may be in the form of
diffuse emission \citep[e.g.][]{Ferguson+96}.  In addition, the calibration
for the SFR derived from H$\alpha$ is only valid for solar metallicity
galaxies, while the outskirts of Arp 78 are likely to have somewhat subsolar
metallicities.  Investigating the metallicity dependence of SFR indicators
with our GALEV models \citep[cf.][]{BickerFritze05}, we found that SFRs in low
metallicity regions are overestimated by factors up to 2 if derived from
L(H$\alpha$) and by up to 50\% if calculated from L(FUV) using calibrations
obtained for close-to-solar metallicity environments.

Even after allowing for these effects, the SFRs derived from the H$\alpha$ and
FUV luminosities are in acceptable concordance. The estimated SFR across the
outer tidal arm structure in Arp~78 as defined in Fig. \ref{fig:haregions} is
$\rm SFR(outer) = 0.2 \pm 0.1\ M_{\odot}\,yr^{-1}$. This result indicates that
the upper mass stellar IMF is normal in the sense that the satisfactory
agreement within reasonable observational uncertainties between SFRs derived
from FUV and H$\alpha$ luminosities does not require significantly abnormal
IMF slope or a peculiarly low upper stellar mass cutoff.

\subsection{Extended UV Emission and the IMF in Arp~78}

The UV-bright outer arm of Arp~78 contains several HII regions with
$\rm L(H\alpha) \geq 10^{38}\,erg\,s^{-1}$.  Photoionization requires some
component of the region to have an age of ${\rm \leq 7-9~Myr}$ and for a
normal mass function the inferred young stellar masses in these regions are
${\rm >10^4\,M_{\sun}}$.  Hence, these young clusters are sufficiently massive
to properly sample the IMF up to high masses
\citep{WeidnerKroupa06,Boissier+07}.

Why then is the XUV outer arm structure so obvious in Arp~78 while the optical
counterpart is faint? This appears to be an effect similar to what is observed
in the context of Tidal Dwarf Galaxies, where the mass and the NIR light can
be dominated by stars inherited from the spiral disk while the
short-wavelength light is due to stars formed in a major outer SF event
triggered by the ongoing interactions in this system. Evidently this tidal arm
feature has not persisted for sufficient time to build up an optically
luminous stellar population. For example, \cite{Neff+05} show that FUV-R
can readily exceed 4 magnitudes in transitory features with ages of $<$
500~Myr.

\section{Summary} 
Deep narrow-band H$\alpha$ imaging from the WIYN 3.5-m telescope on Kitt Peak
allowed us to detect H$\alpha$ emission in spatial coincidence with the
extended GALEX FUV structure across the inner part and, in particular, across
the XUV extended outer structures of Arp 78, that very probably are of tidal
origin. We calculated the SFR across these outer structures and found
  a value of $\rm 0.2 \pm 0.1\ M_{\odot}\,yr^{-1}$ from both the FUV and
  H$\alpha$ luminosities. The agreement between these two measures points to
  a relatively normal upper stellar IMF in this system. We also find the
  individual star forming regions in this system to be younger than 15 Myr and
  sufficiently massive to properly sample the IMF up to high masses.

The XUV-strong and optically faint outer structures of Arp~78 are consistent
with ongoing SF superimposed on an older stellar population torn out from the
spiral disk by tidal forces. A detailed investigation of stellar population
ages and metallicities across Arp~78 will be the subject of a forthcoming
paper.

\acknowledgments
These data were obtained as a result of support of the WIYN Observatory by the
University of Wisconsin-Madison. We thank our referee, Dr. Gerhardt Meurer,
for his very insightful and helpful comments which helped to improve and
clarify this paper. We are also grateful to the WIYN staff for their many
contributions to the success of these WIYN observations. JSG gratefully
acknowledges partial support for this research from the National Science
Foundation through grant AST-0708967 and from the University of Wisconsin
Graduate School.



{\it Facilities:} \facility{WIYN (MiniMo)}, \facility{GALEX ()}.

\bibliographystyle{bibstyle}
\bibliography{ms}

\end{document}